\documentclass[letterpaper]{article} % DO NOT CHANGE THIS
\usepackage[draft]{aaai25}  % DO NOT CHANGE THIS
\usepackage{times}  % DO NOT CHANGE THIS
\usepackage{helvet}  % DO NOT CHANGE THIS
\usepackage{courier}  % DO NOT CHANGE THIS
\usepackage[hyphens]{url}  % DO NOT CHANGE THIS
\usepackage{graphicx} % DO NOT CHANGE THIS
\usepackage{subcaption}  % %%%% Matti added this for image subcaptions
\usepackage[export]{adjustbox}   % %%%%% Matti added this for frame around screen shots
\usepackage{lipsum} % %%%%% Matti added this just for laughs

\usepackage{natbib}  % DO NOT CHANGE THIS AND DO NOT ADD ANY OPTIONS TO IT
\usepackage{caption} % DO NOT CHANGE THIS AND DO NOT ADD ANY OPTIONS TO IT
\usepackage{algorithm}
\usepackage{algorithmic}

\usepackage{newfloat}
\usepackage{listings}

\DeclareCaptionStyle{ruled}{labelfont=normalfont,labelsep=colon,strut=off} % DO NOT CHANGE THIS
\floatstyle{ruled}
\newfloat{listing}{tb}{lst}{}
\floatname{listing}{Listing}
\title{An XAI Social Media Platform for Teaching K-12 Students AI-Driven Profiling, Clustering, and Engagement-Based Recommending}
\author{
    Nicolas Pope\textsuperscript{\rm 1},
    Juho Kahila\textsuperscript{\rm 2},
    Henriikka Vartiainen\textsuperscript{\rm 2},
    Mohammed Saqr\textsuperscript{\rm 1},
    Sonsoles López-Pernas\textsuperscript{\rm 1},
    Teemu Roos\textsuperscript{\rm 3},
    Jari Laru\textsuperscript{\rm 4},
    Matti Tedre\textsuperscript{\rm 1}
}
\affiliations{
    \textsuperscript{\rm 1}University of Eastern Finland, School of Computing, Joensuu, Finland\\
    \textsuperscript{\rm 2}University of Eastern Finland, Applied Educational Science and Teacher Education, Finland\\
    \textsuperscript{\rm 3}University of Helsinki, Department of Computer Science, Helsinki, Finland\\
    \textsuperscript{\rm 4}University of Oulu, Faculty of Educational Sciences and Psychology, Oulu, Finland\\
    \{npope, juho.kahila, henriikka.vartiainen, mohammed.saqr, sonsoles.lopez\}@uef.fi, teemu.roos@helsinki.fi, jari.laru@oulu.fi, matti.tedre@uef.fi
}

\begin{document}

\maketitle

\begin{abstract}
This paper, submitted to the special track on resources for teaching AI in K-12, presents an explainable AI (XAI) education tool designed for K-12 classrooms, particularly for students in grades 4-9.  The tool was designed for interventions on the fundamental processes behind social media platforms,  focusing on four AI- and data-driven core concepts: data collection, user profiling, engagement metrics, and recommendation algorithms.  An Instagram-like interface and a monitoring tool for explaining the data-driven processes make these complex ideas accessible and engaging for young learners.  The tool provides hands-on experiments and real-time visualizations, illustrating how user actions influence both their personal experience on the platform and the experience of others.  This approach seeks to enhance learners' data agency, AI literacy, and sensitivity to AI ethics.  The paper includes a case example from 12 two-hour test sessions involving 209 children, using learning analytics to demonstrate how they navigated their social media feeds and the browsing patterns that emerged.
\end{abstract}


% This special track invites papers on the development and use of resources to support K-12 AI education. Examples include online demos, software tools, and structured activities. Submissions should follow the standard EAAI format for an academic paper and include the following: description of the resource; target age group; setup and resources needed; AI concepts addressed; expected learning outcomes; and (if possible) implementation results. Online demos and software tools should be accompanied by brief video walk-throughs.

\section{Introduction}

Over the past fifteen years, social media has evolved from an emerging Internet trend into a dominant societal force that shapes cultural trends, public discourses,  communication practices, and political landscapes \cite{ortiz19}.  Teenagers and younger children are a quickly growing user group: A 2022 survey showed that 95\% of teens aged 13 to 17 used social media, with platforms like TikTok (67\%), Instagram (62\%), and Snapchat (59\%) being especially popular \cite{pew22}. Although platforms typically set a minimum age to 13, also many younger children are active users \cite{vartiainen23c}.  For young people, social media provides an important connection to peers and a platform for creativity and informal learning---but it also exposes them to risks like mis- and disinformation, loss of privacy, and cyberbullying \cite{vartiainen23c,hendricks19,ng22,ito23}.  Engaging with social media platforms influences their social lives, cognitive development, attitudes, and self-esteem \cite{mccosker17,valkenburg22}.  

The growing engagement of youth with social media raises questions about how to support young learners' data agency---their ability and volition to take informed actions that impact their digital worlds---and equip them with the skills needed to navigate those worlds safely and responsibly \cite{vartiainen22}.  The concerns with the impact of social media have fueled a quickly growing body of research on social media literacy.  Much of this research has focused on teaching novice users how to protect their privacy, recognize harmful behaviors, and critically evaluate content  \cite{stoilova20,livingstone14,selwyn18,pangrazio19,keen22,swart21}.   Educational interventions for mitigating the risks of social media use have included guidelines for identifying fake news and social media literacy education into the classroom \cite{civek18,hobbs18,mason18,moore22}.

However, many educational initiatives focus on the uses and effects of social media, but do not address the underlying data-driven mechanisms that power those platforms.  While youth are active on social media, they often use the platforms unaware of the extent, types, and uses of data collected about them, as well as the mechanisms for collecting and using the data \cite{pangrazio19,vartiainen22,swart21}. As a result, users base their data tactics and strategies on ``folk theories''---intuitive, informal, yet often incorrect beliefs---to explain how social media platforms work \cite{eslami16,devito17}, which can lead to a false sense of security and control.  At the same time platforms increasingly adopt new data-driven approaches for marketing, behavior engineering, and opinion shaping \cite{zuboff19,fisher22}.  

Social media platforms represent one of the most tangible and immediate interactions youth have with AI and data-driven technology.  Although children and youth may recognize manipulation attempts \cite{vartiainen23c}, their ability to mitigate risks is reduced by limited understanding of how sophisticated algorithms track user behavior, build detailed profiles, and deliver personalized content \cite{schmeichel18,cherner19,mertala21b,vartiainen22}.  Research has shown that schools play a crucial role: just using social media platforms does not develop an accurate understanding of how they actually work \cite{mertala21b,vartiainen22}, (Authors, 2024).  While social media literacy frameworks have noted the importance of technical competencies and digital literacies \cite{cho24,schreurs21}, a recent review suggests that the field often prioritizes use skills over an understanding of how the systems work \cite{polanco22}.

\begin{figure}[]
\centering
\includegraphics[width=0.7\columnwidth, frame]{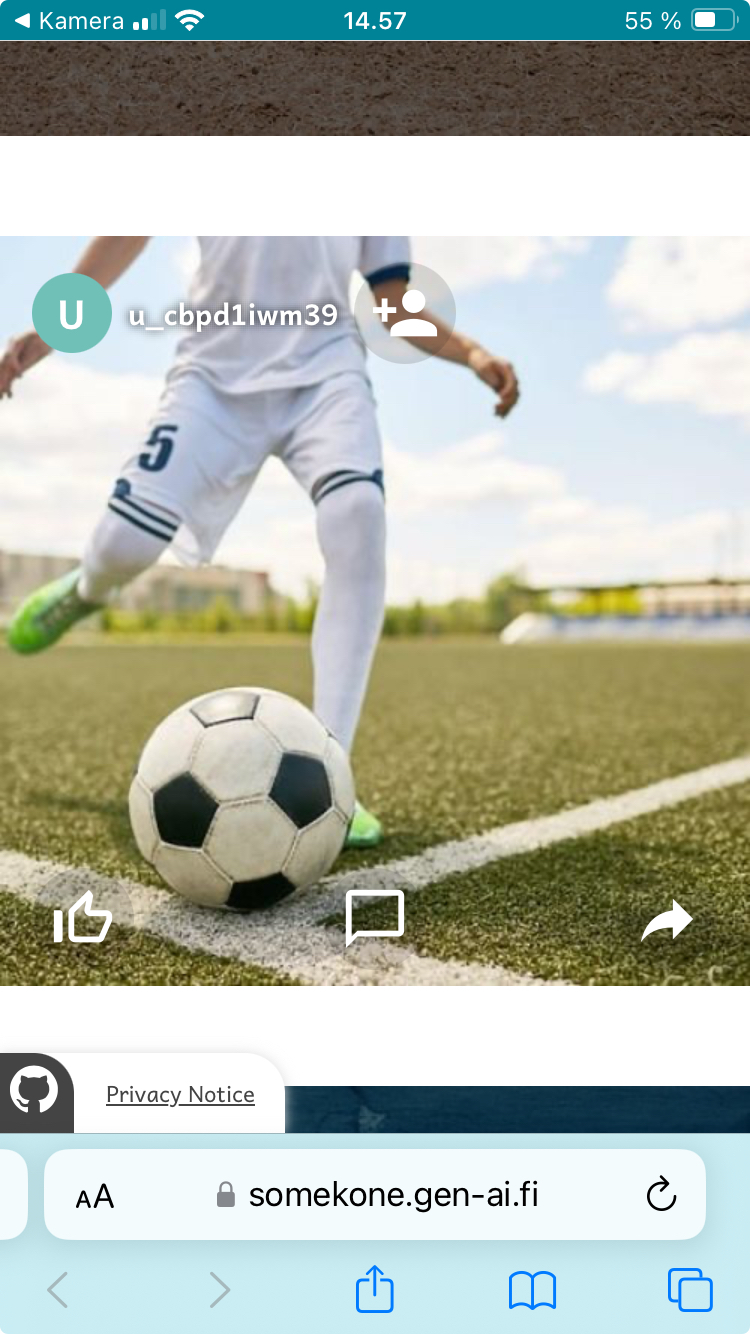} 
\caption{Browsing an image feed on Somekone.}
\label{fig:app_imagefeed}
\end{figure}

A number of AI education tools address aspects of social media, such as filter bubbles \cite{janssen21}, recommendations \cite{tsiakas20,taibi21}, and general platform functionalities \cite{nagulendra13,hartl24,difranzo19}.  A number of tools that are not empirically studied exist with similar range of aims (see a survey by Authors, forthcoming).  However, these tools typically lack a comprehensive approach that connects individual user's actions to the personal and societal impacts of AI and data-driven systems.  To address that gap, this paper presents an Instagram-like explainable AI (XAI) tool for learning about the mechanisms of data collection, engagement, profiling, and recommendation on social media platforms.  The paper also shows how the tool can be applied in social network analysis and learning analytics research.

% %%%%%%%%%%%%%%%%%%%%%%%%%%%%%%%%%%%%%%%%%%%%%%%%%%%%%%%%%%%%%%%%%%%%%%%%%%
% %%%%%%%%%%%%%%%%%%%%%%%%%%%%%%%%%%%%%%%%%%%%%%%%%%%%%%%%%%%%%%%%%%%%%%%%%%
% %%%%%%%%%%%%%%%%%%%%%%%%%%%%%%%%%%%%%%%%%%%%%%%%%%%%%%%%%%%%%%%%%%%%%%%%%%

\section{Intended Learning Outcomes}

\paragraph{Target age groups.}
The AI education tool ``Somekone'' is designed for use in classrooms with students aged 11-16 years (grades 4-9), but can also be used in pre-service and in-service teacher education (Authors, forthcoming).  It features an Instagram-like interface, enabling students to learn social media mechanisms in an accessible environment familiar to many in that age group \cite{vartiainen23c,ortiz19} (See Fig. \ref{fig:app_imagefeed}).   %SAuthors=SIGCHI PAPER

\paragraph{AI concepts addressed.}
The intended learning outcomes (ILOs) focus on four core concepts related to AI and data-driven automation: data collection, profiling, engagement, and recommendation.  Each of those high-level concepts invokes a number of other, more specific concepts and they align with several ``Five Big Ideas in AI,'' especially representation and reasoning, learning, natural interaction, and societal impact \cite{touretzky23}.  Through classroom activities and teacher facilitation, students gain a deeper understanding of how social media platforms operate and how their own behavior shapes and is shaped by these technologies.  A detailed report of the results and guidelines on implementing these classroom activities is provided elsewhere (Authors, forthcoming).

\emph{Data Collection.}
Students explore how social media platforms collect various types of data, both  overtly and covertly.  This includes \textit{data given}, or information explicitly entered by users, such as username and comments; \textit{data traces}, or data observed from user interactions, such as likes or time spent viewing content; and \textit{data inferred}, or information derived by combining user data with that of others, such as image preferences \cite[for discussion of the importance of these see][]{vanderhof17,pangrazio19,hendricks19}.  Understanding these data types is necessary for appreciating their privacy and security implications \cite{barassi20} as well as for understanding how profiling works, particularly since much of this data collection is not visible to users.

\emph{Profiling.}
Students learn how social media platforms profile users based on their behavior and interests. The underlying AI concepts include vector-based data representations, clustering, and network analysis.  Students also learn that profiling processes, though often inaccurate and biased \cite{noble18,benjamin19,morales24}, are used to personalize online experiences, including their uses in manipulation and targeted advertising \cite{eubanks18,zuboff19}. In the classroom, the tool enables students to examine the ethical implications of profiling.

\emph{Engagement.}
Students investigate how social media platforms track interactions and data given (e.g., likes, comments) as well as data traces (e.g., time spent viewing content) to increase user retention. They learn how engagement metrics are calculated and used to keep users on the platform longer, often at the expense of their well-being \cite{valkenburg22}.  The tool illustrates engagement metrics that are typically hidden from users, connecting the concept to the mechanisms of recommendation systems.

\def\minipicturesize{0.9} % Size of multi-figure subfigures
\def\miniminipicturesize{0.235} % Size of multi-figure subfigures
\def\miniminiscreenshotsize{0.705} % Size of large subfigures
\def\miniminiteacherviewsize{0.325} % Size of medium subfigures

\begin{figure*}[]
    \centering

    \begin{subfigure}[t]{\miniminipicturesize\textwidth}
        \centering
        \captionsetup{justification=centering}
        \includegraphics[height=220px, frame]{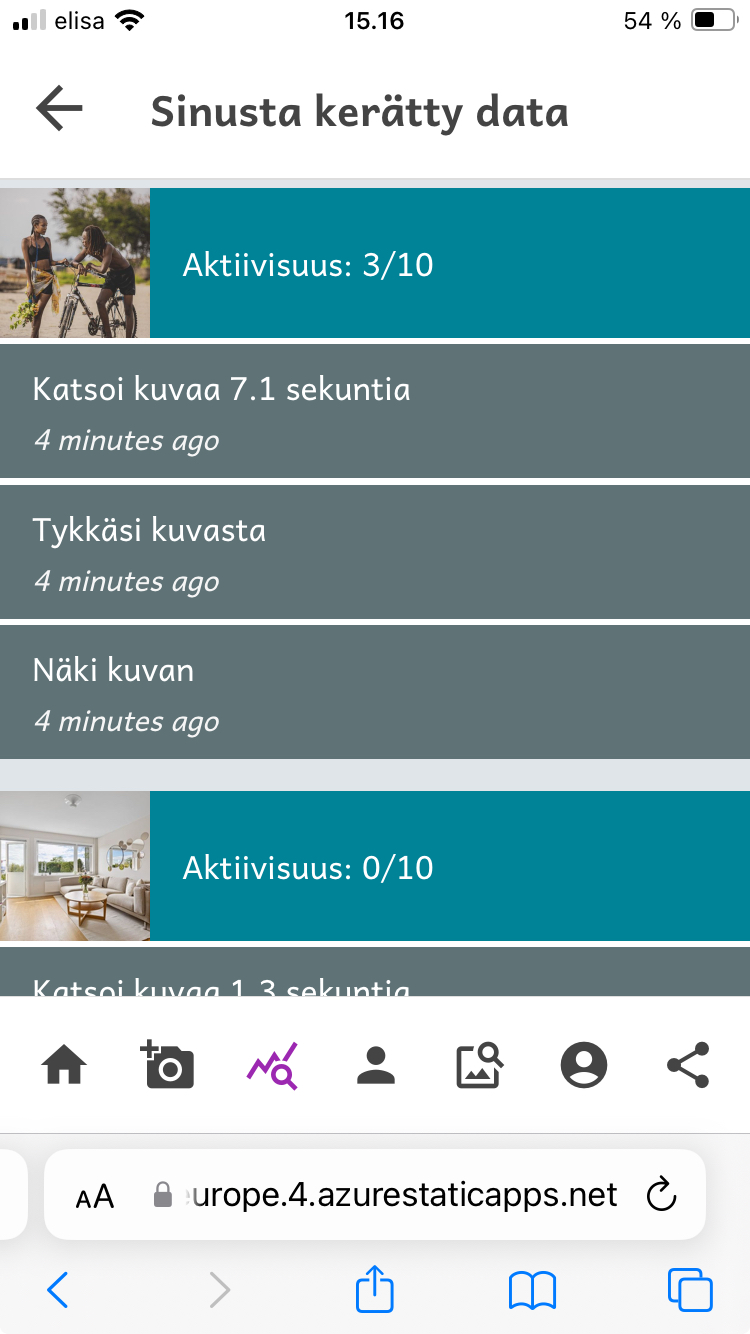}
        \caption{Live view of the data captured from a user's browsing.}
        \label{fig:app_dataview}
    \end{subfigure}
    ~ 
    \begin{subfigure}[t]{\miniminiscreenshotsize\textwidth}
        \centering
        \captionsetup{justification=centering}
        \includegraphics[height=220px, frame]{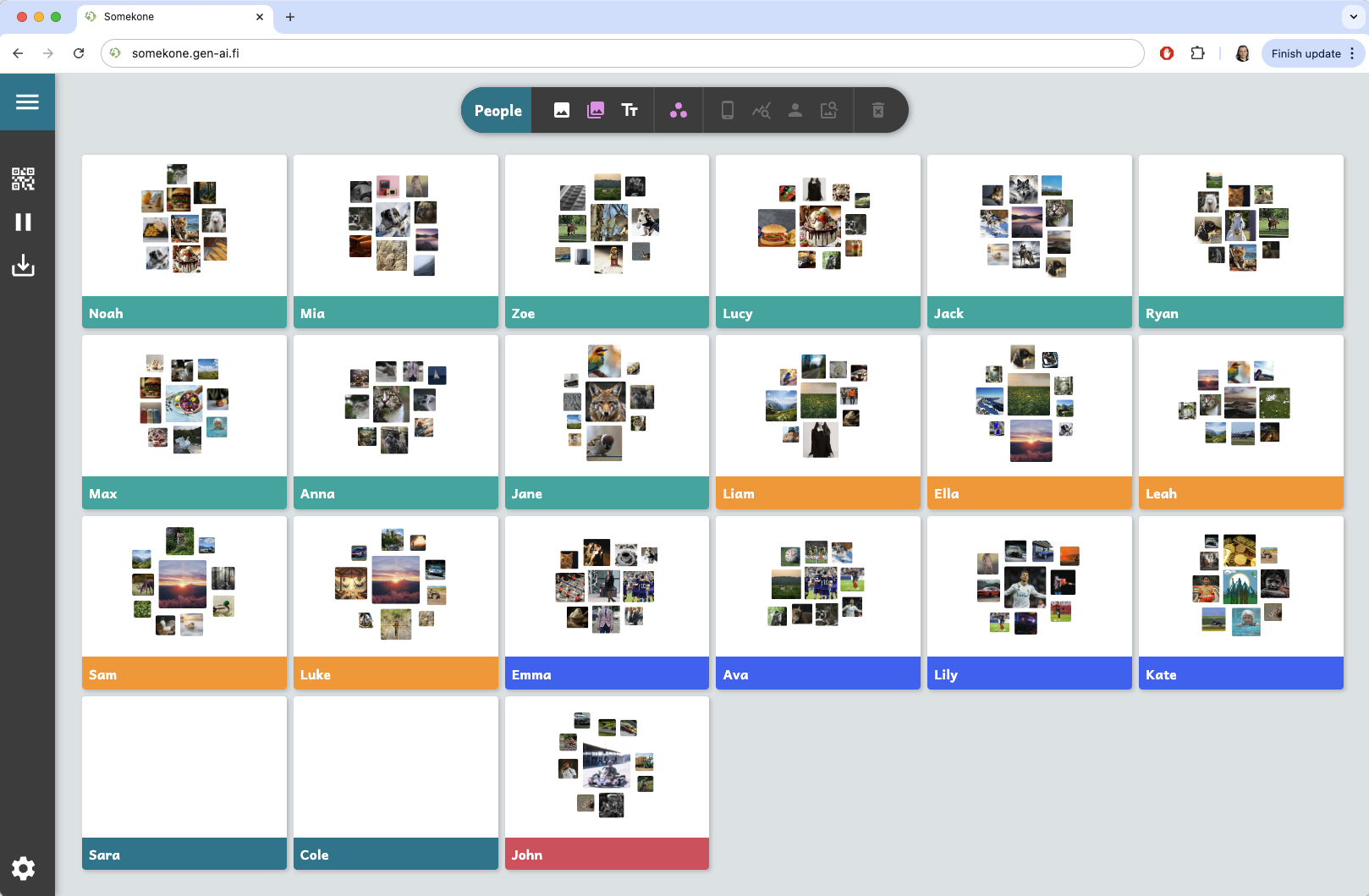}
        \caption{A view of all users' most engaged images, shown on the classroom projector.}
        \label{fig:cls_engaged}
    \end{subfigure}
    \caption{Somekone users can work in pairs, where one user browses the image feed on a mobile device and another connects a second device to that feed to analyze a live view of data collection (Fig.\ \ref{fig:app_dataview}).  At the same time, the whole classroom's most engaged images are shown on the classroom projector (Fig.\ \ref{fig:cls_engaged}).}
\end{figure*}

\emph{Recommending.}
Students learn how social media platforms recommend content based on user profiles and their similarities with other users. The relevant AI concepts include similarity search, supervised learning, and learning to rank techniques.  Designed to boost user engagement, the recommendation processes are typically opaque, making it difficult for users to understand why they are shown certain content.  The tool introduces students to a number of basic recommendation mechanisms and lets users experiment with them, such as personalized vs.\ global recommendations; \#hashtag vs.\ user similarity vs.\ co-engagement based suggestions; optimized vs.\ randomized; and the impact of diversity on recommendations.  

Beyond these outcomes, the tool can support broader learning objectives linked with classroom activities.  For example, the personal and societal impacts of social media can be explored in the classroom, including how these platforms influence opinions, behaviors, social interactions, and societal trends (Authors, forthcoming).  Classroom activities can focus on, for instance, marketing, spread of misinformation, and polarization (ibid.), helping students to reflect on their own experiences and guide them toward informed actions and decisions online.

% %%%%%%%%%%%%%%%%%%%%%%%%%%%%%%%%%%%%%%%%%%%%%%%%%%%%%%%%%%%%%%%%%%%%%%%%%%
% %%%%%%%%%%%%%%%%%%%%%%%%%%%%%%%%%%%%%%%%%%%%%%%%%%%%%%%%%%%%%%%%%%%%%%%%%%
% %%%%%%%%%%%%%%%%%%%%%%%%%%%%%%%%%%%%%%%%%%%%%%%%%%%%%%%%%%%%%%%%%%%%%%%%%%

\section{Description of the Resource}

% Nick: We are illustrating an application of AI as intelligent decision making that directly influences them. From an AI concept perspective it is data + learning + metrics that are targeted by the learning (engagement) and the actions which the AI can take, ie. using the output to choose content. It is similar to the TM except they encounter different kinds of data and a continuous score rather than discrete classifications. They are also exposed to more of the blackbox steps that are hidden in a neural network, for instance clustering, principle components (kinda, the topic affinities). It is also a hybrid of algorithmic and ML.

\begin{figure*}[!t]
    \centering

    \begin{subfigure}[t]{\miniminipicturesize\textwidth}
        \centering
        \captionsetup{justification=centering}
        \includegraphics[width=\minipicturesize\linewidth, frame]{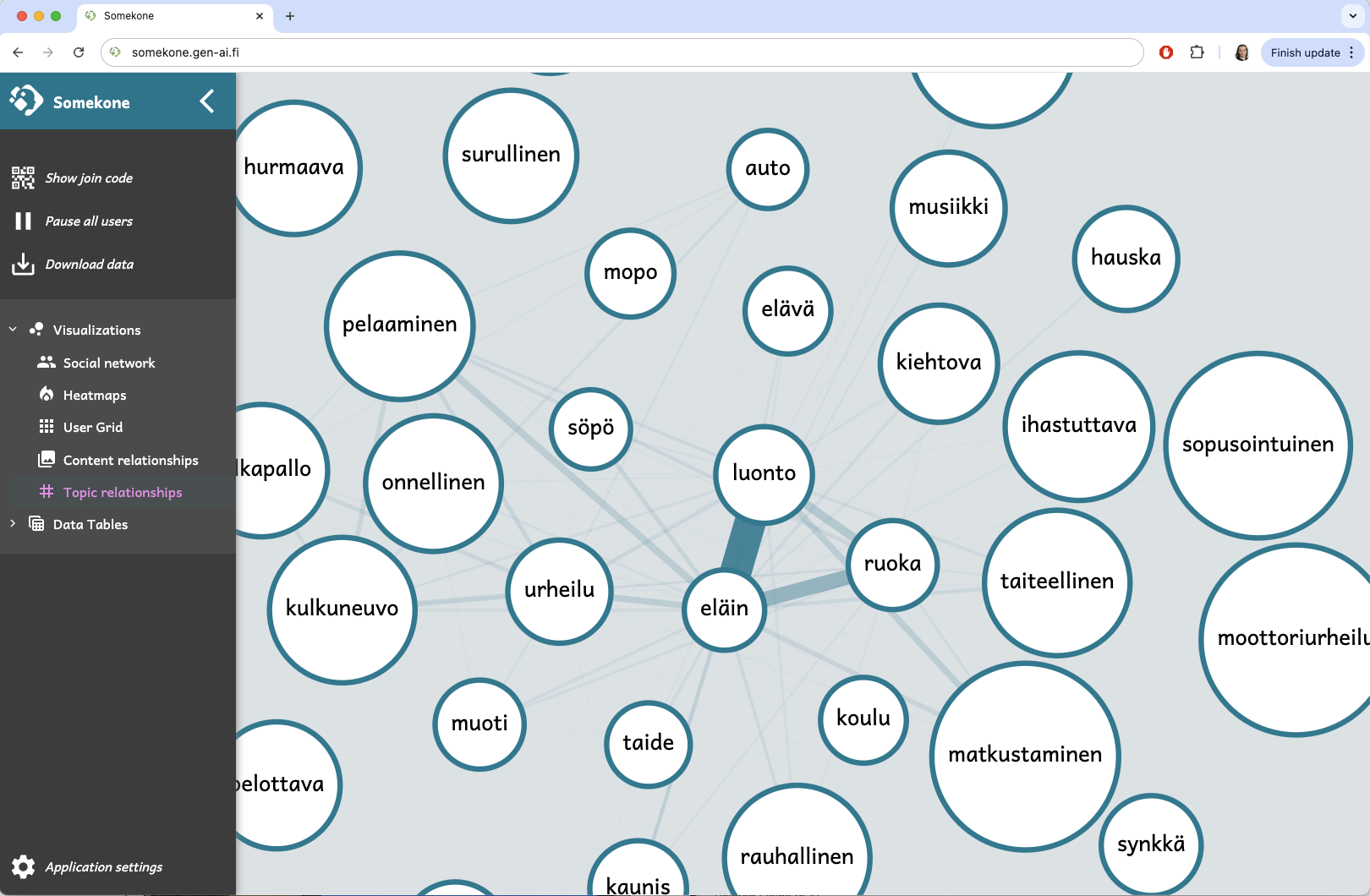}
        \caption{Topic affinities}
        \label{fig:scr_topic_affinity}
    \end{subfigure}
    ~ 
    \begin{subfigure}[t]{\miniminipicturesize\textwidth}
        \centering
        \captionsetup{justification=centering}
        \includegraphics[width=\minipicturesize\textwidth, frame]{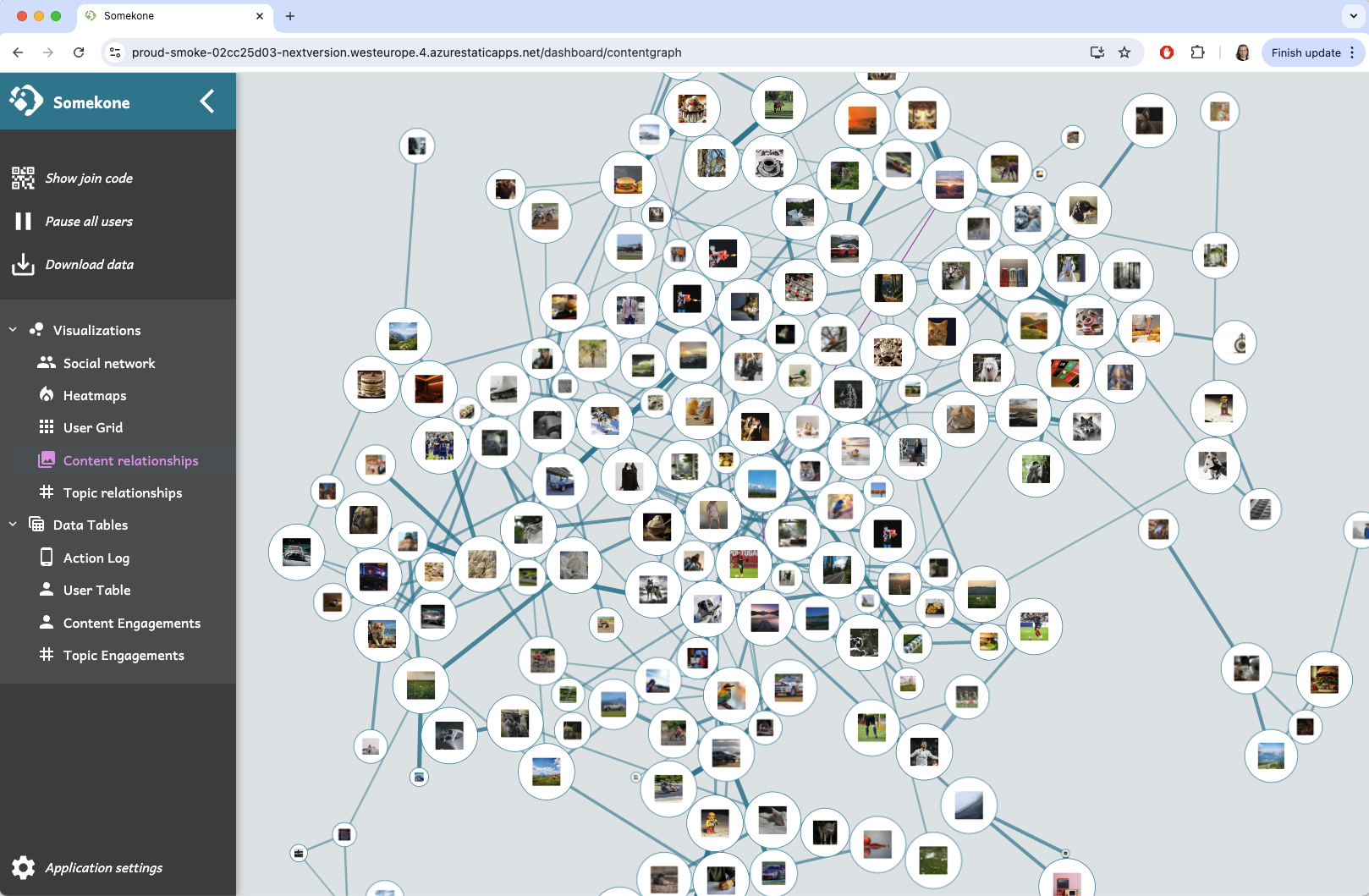}
        \caption{Image coengagement.}
        \label{fig:scr_image_coengagement}
    \end{subfigure}%
    ~ 
    \begin{subfigure}[t]{\miniminipicturesize\textwidth}
        \centering
        \captionsetup{justification=centering}
        \includegraphics[width=\minipicturesize\linewidth, frame]{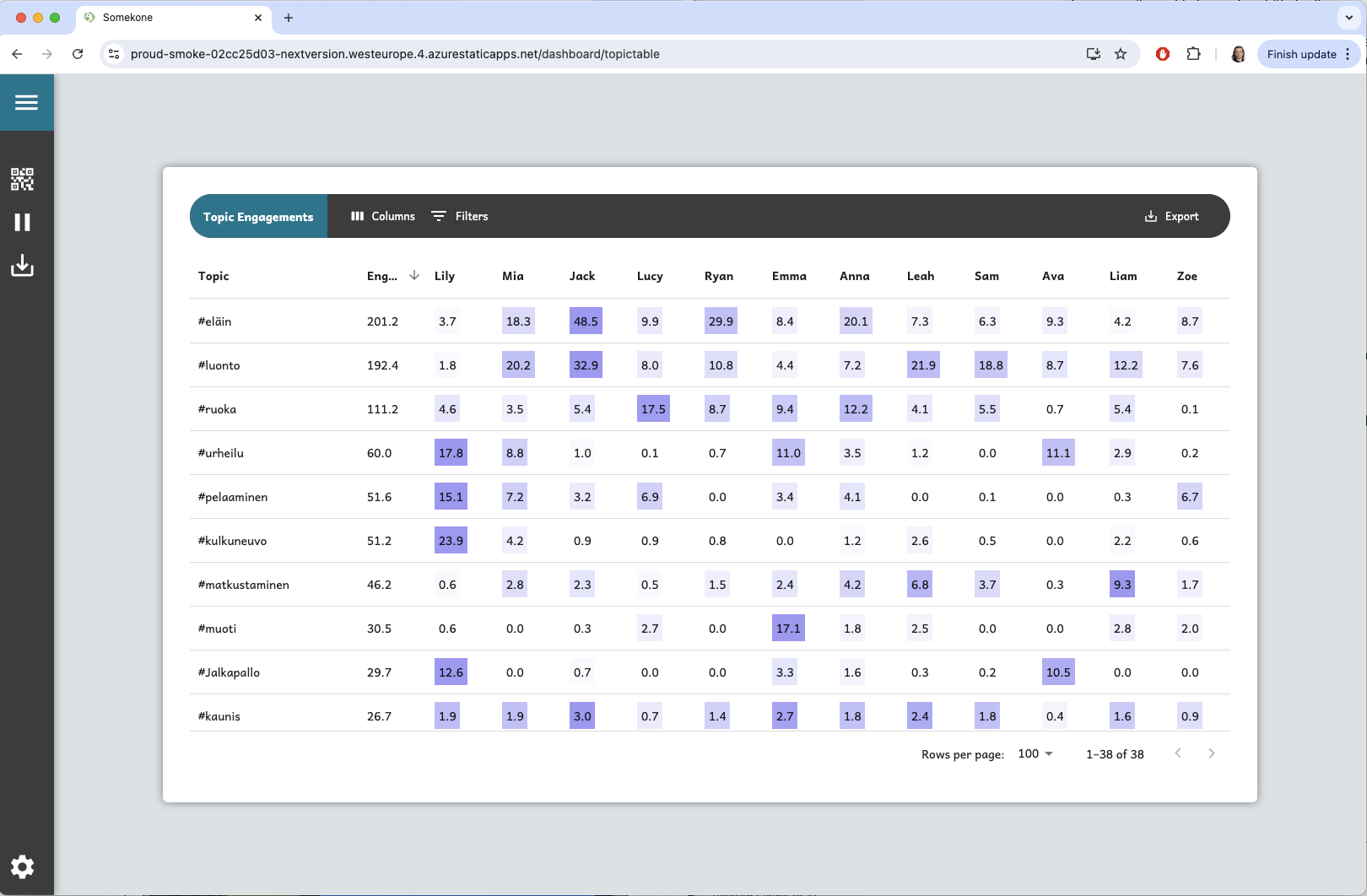}
        \caption{Table of users' affinities.}
        \label{fig:scr_table_view}
    \end{subfigure}%
    ~ 
    \begin{subfigure}[t]{\miniminipicturesize\textwidth}
        \centering
        \captionsetup{justification=centering}
        \includegraphics[width=\minipicturesize\linewidth, frame]{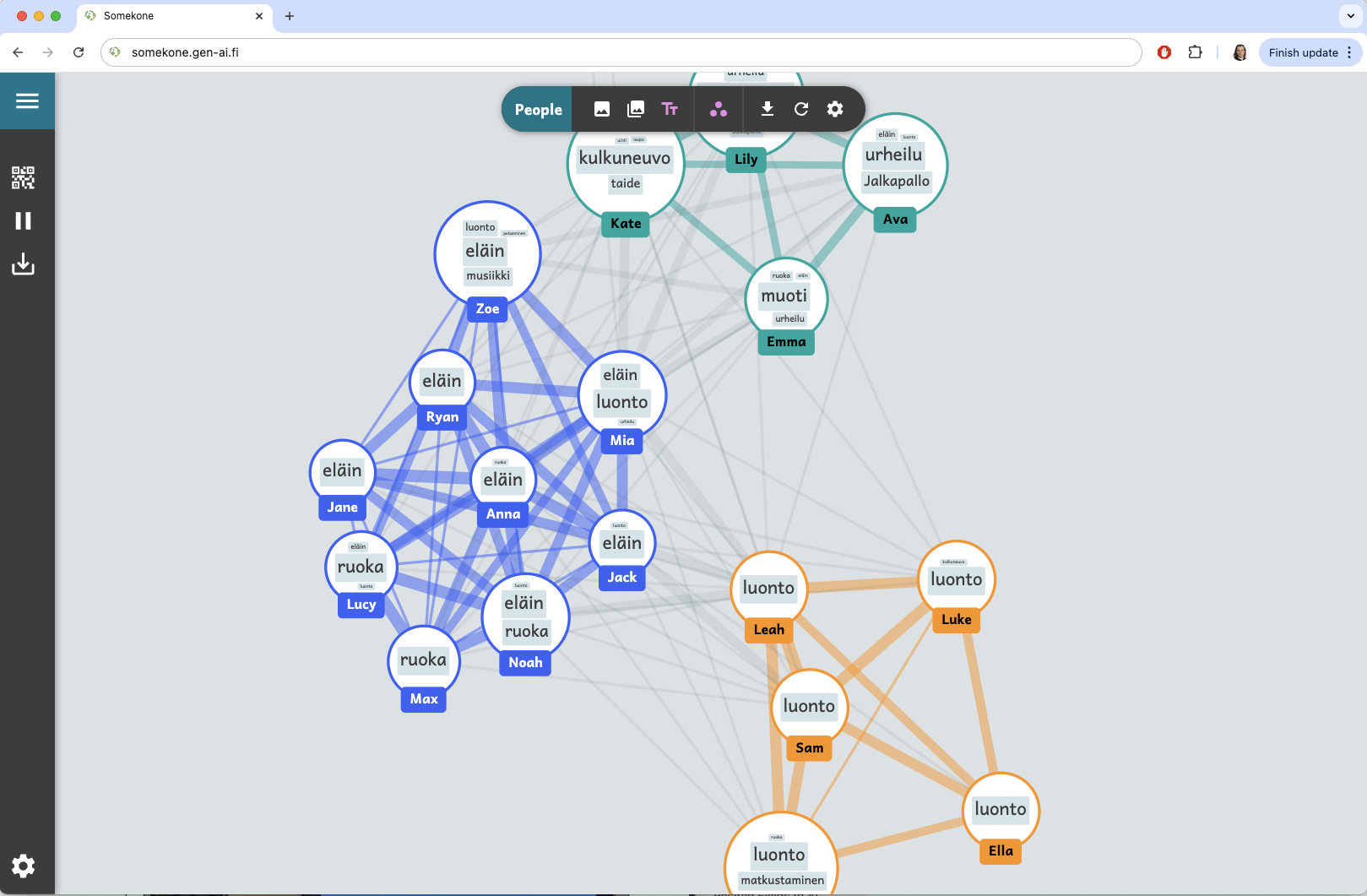}
        \caption{Profile clustering.}
        \label{fig:scr_topic_clustering}
    \end{subfigure}
    \caption{Example views on teacher screen.}
\end{figure*}

The Somekone tool is designed to illustrate key AI concepts, including (i) data collection, (ii) profiling, (iii) engagement, and (iv) content recommendation.  In addition to individual views on students' devices, Somekone offer a teacher's view on the classroom projector, which can also illustrate (v) clustering through visualizations of clustered and color-coded social networks, (vi) co-engagement by showing networks of images liked by the same users (Fig. \ref{fig:scr_image_coengagement}), (vii) topic affinity by grouping \#hashtags liked by the same users (Fig. \ref{fig:scr_topic_affinity}), and (viii) recommendation algorithms through heat maps that visualize how different recommendation algorithms impact what content a user is likely to be shown and what content the user is likely \emph{not} shown.

\paragraph{The image feed view.}
Somekone features an Instagram-like interface with an infinite scrolling image feed as well as basic social media functions such as liking, reacting with emojis, commenting, following, and sharing (Fig. \ref{fig:app_imagefeed}).  The default content set includes 727 images, selected and labeled by two classes of eight-graders, with further curation by two researchers.  During typical classroom sessions, students work in pairs, with one device for browsing the image feed.

\paragraph{The data collection view.}
Somekone supports collaborative learning by allowing one device to browse the image feed while another device displays an analytics dashboard on that browsing session.  Alternatively, a single user can swap between image feed and analytics views on one device.  The browsing device tracks user interactions, such as time spent on each image, likes, comments, comment length, reactions, following, and periods of inactivity. These interactions are recorded into an action log. The paired device provides a real-time view of the action log, allowing the students to observe the immediate effects of their browsing.  Figure \ref{fig:app_dataview} shows an example data collection view (user watched the image for 7.1 seconds and clicked on ``like'') and an engagement score 3/10 calculated from those data points.  The teacher's computer can project a classroom-wide view of all students in the classroom and their most engaged images (Fig. \ref{fig:cls_engaged}) (classroom views are configurable, for instance engagement can be shown as an image cloud or heatmap).

\begin{figure*}[]
    \centering

    \begin{subfigure}[t]{\miniminipicturesize\textwidth}
        \centering
        \captionsetup{justification=centering}
        \includegraphics[height=220px, frame]{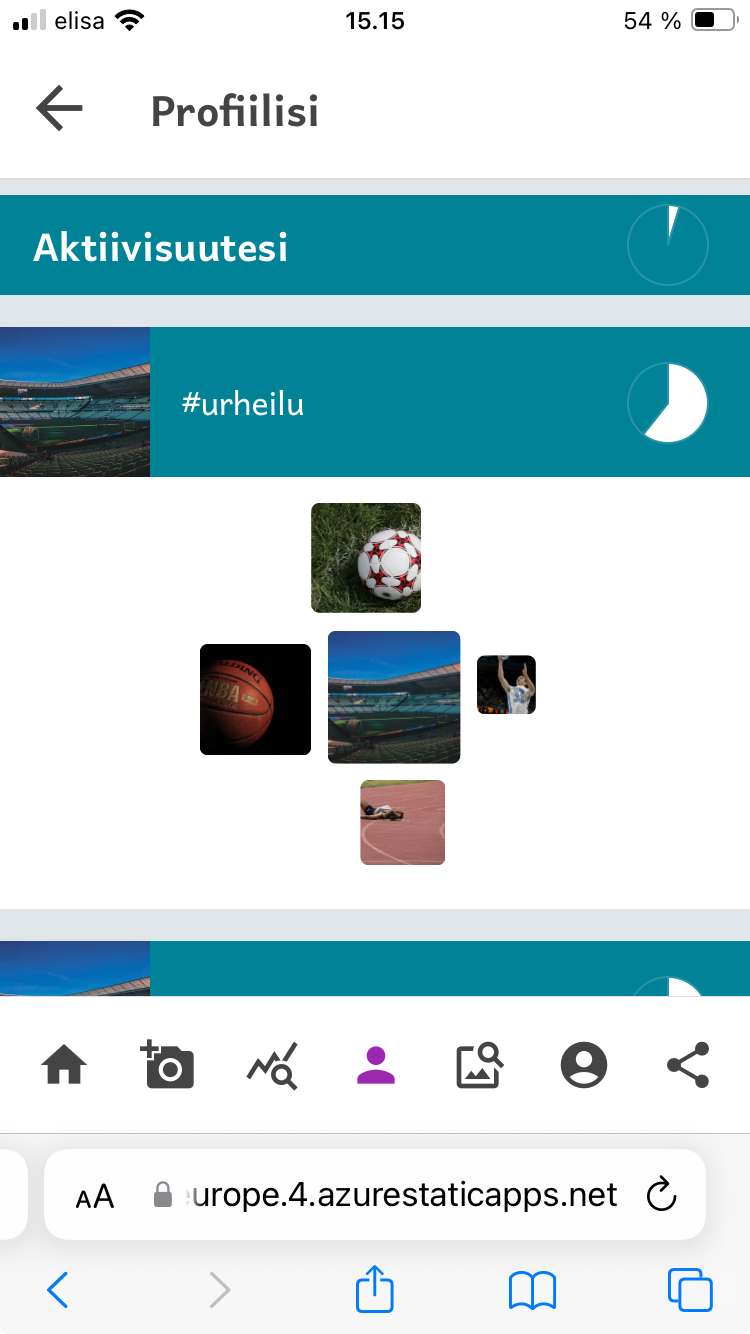}
        \caption{Profile created of browsing data, updated real-time.}
        \label{fig:app_profile}
    \end{subfigure}
    ~ 
    \begin{subfigure}[t]{\miniminiscreenshotsize\textwidth}
        \centering
        \captionsetup{justification=centering}
        \includegraphics[height=220px, frame]{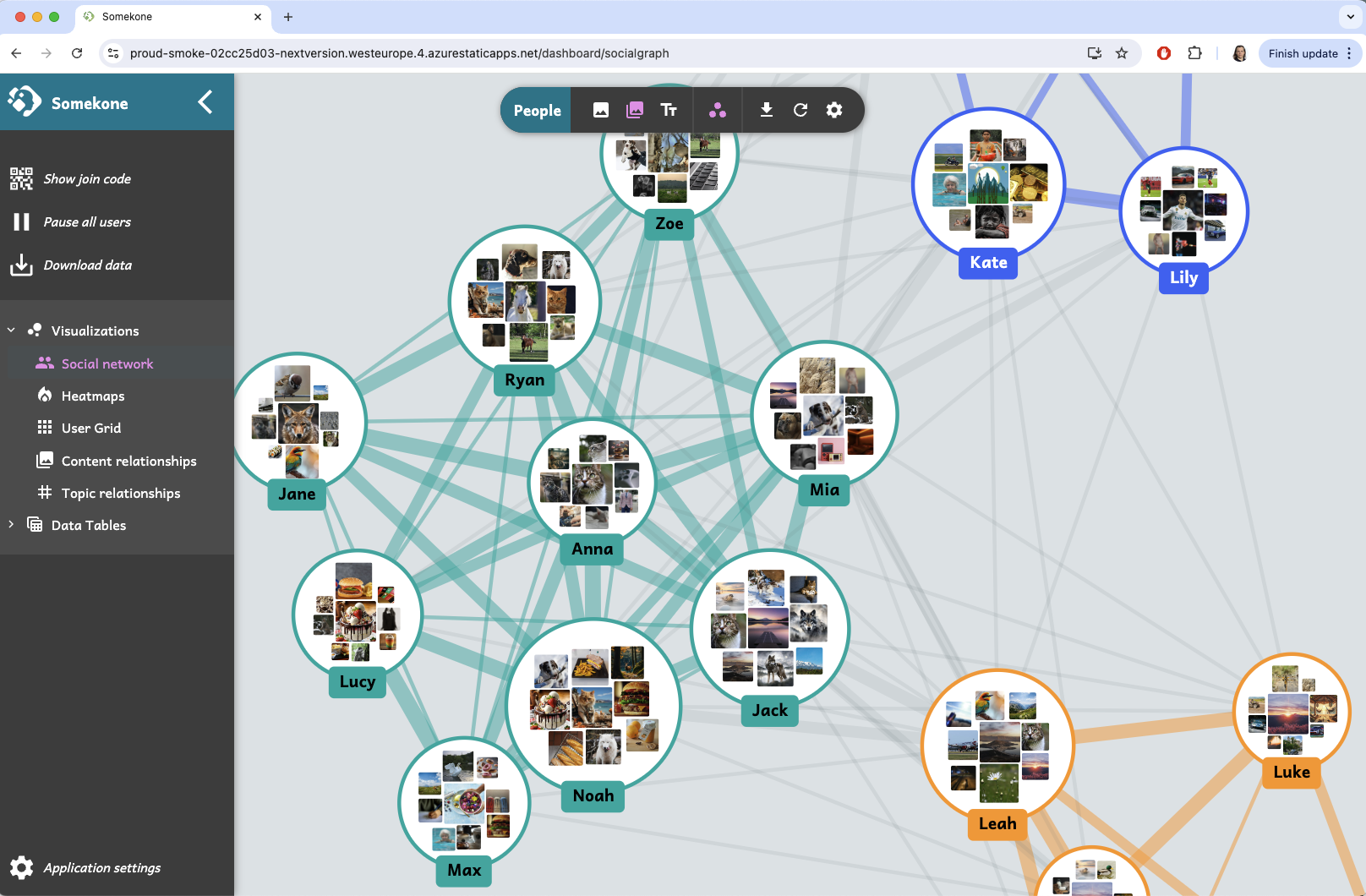}
        \caption{Live view of the classroom's social network, visualizing each users' most engaged pictures, clustered by profile similarity, and shown on the classroom projector.}
        \label{fig:cls_social_network}
    \end{subfigure}
    \caption{A paired user device can show a real-time view of another user's profile forming as they browse the feed, with a breakdown that explains the profile (Fig.\ \ref{fig:app_profile}).  The teacher's view shown on the classroom projector has dozens of visualization options into the classroom social network (Fig.\ \ref{fig:cls_social_network})
    }
\end{figure*}

\paragraph{The profile view.}
In profile view mode, the paired device(s) shows a real-time view of the user's profile based on their engagement with images tagged with specific topic \#hashtags.  The profile view shows these tags in a word cloud, visualizing the user's strongest topic affinities and explains the main user activities that underlie those affinities. Figure \ref{fig:app_profile} shows how a user's profile is built and visualized, and Somekone updates this profile in real-time as browsing continues.  Figure \ref{fig:cls_social_network} shows the classroom view, including clusters of similar profiles, strength of similarity between profiles (thickness of connecting lines), and the most engaged images from each user.

\begin{figure*}[!t]
    \centering

    \begin{subfigure}[t]{\miniminipicturesize\textwidth}
        \centering
        \captionsetup{justification=centering}
        \includegraphics[height=220px, frame]{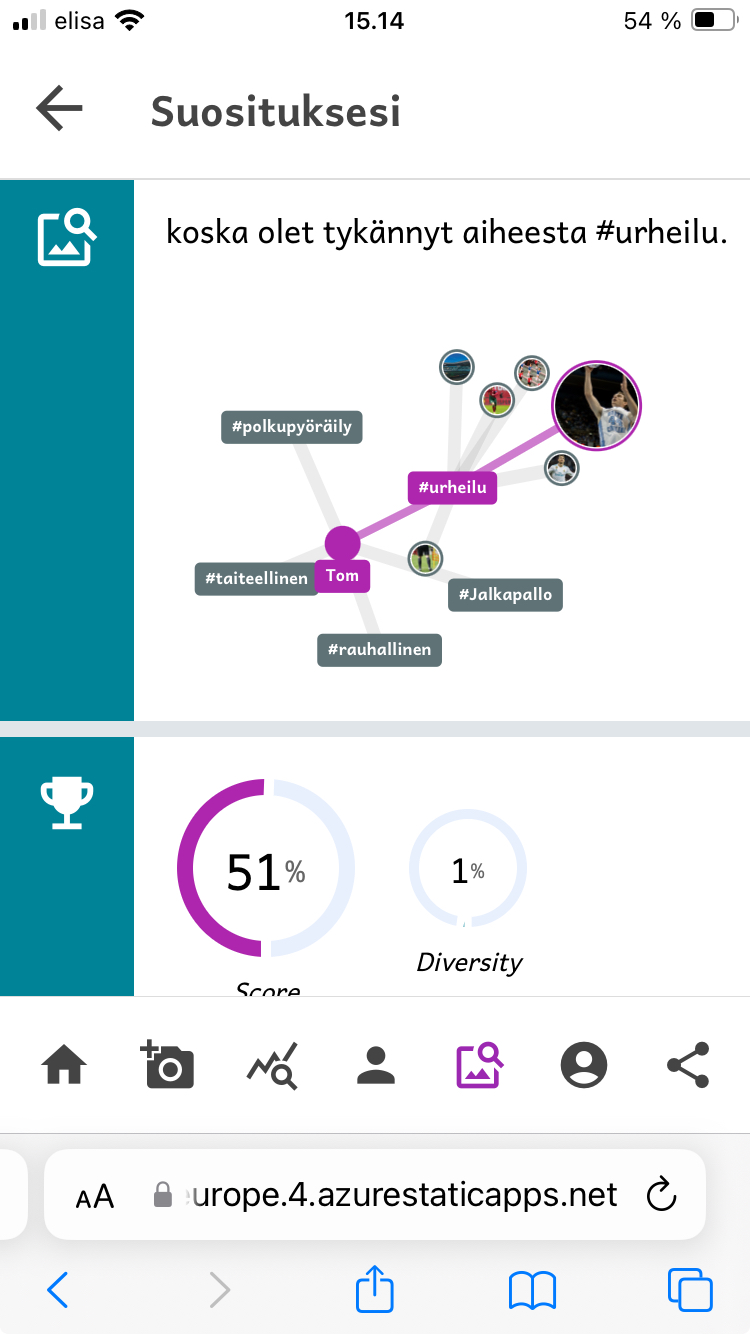}
        \caption{Live XAI view of upcoming recommendations.}
        \label{fig:app_recommendations}
    \end{subfigure}
    ~ 
    \begin{subfigure}[t]{\miniminiscreenshotsize\textwidth}
        \centering
        \captionsetup{justification=centering}
        \includegraphics[height=220px, frame]{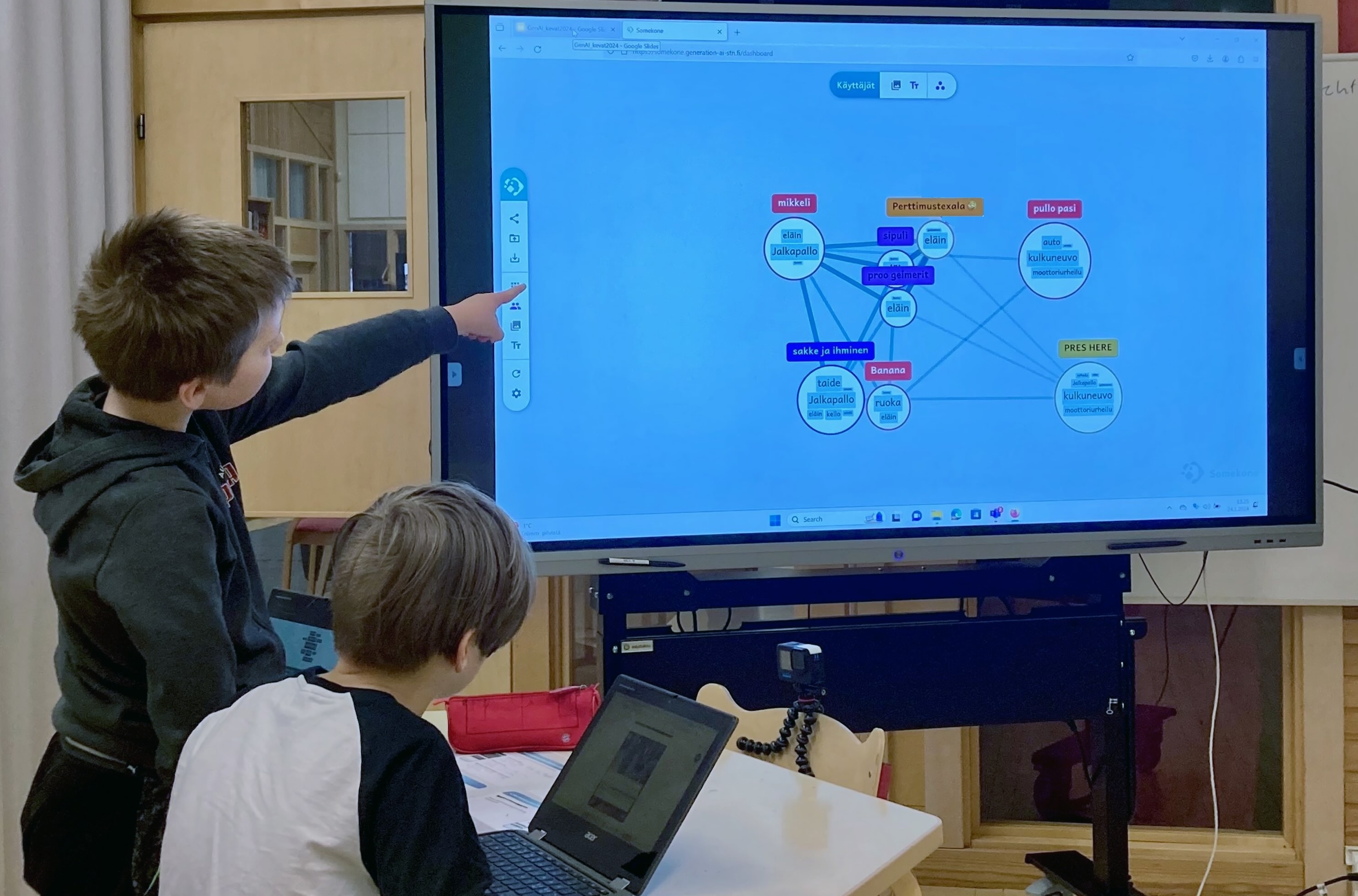}
        \caption{The classroom projector showing a real-time visualization of all learners' profiles as \#hashtag clouds, clustered by profile similarity, and students analyzing the nearest profiles in the class.}
        \label{fig:cls_clustering}
    \end{subfigure}
    \caption{The recommendations view of Somekone shows, in advance, the next content recommended to the paired user, with a detailed breakdown for each recommendation (Fig.\ \ref{fig:app_recommendations}).  The classroom view can provide different visualizations of the clusters that serve as a basis for recommendations (Fig.\ \ref{fig:cls_clustering}.)
    }
\end{figure*}

\begin{figure*}[!t]
     \centering

    \begin{subfigure}[t]{\miniminiteacherviewsize\textwidth}
        \centering
        \captionsetup{justification=centering}
        \includegraphics[height=200px]{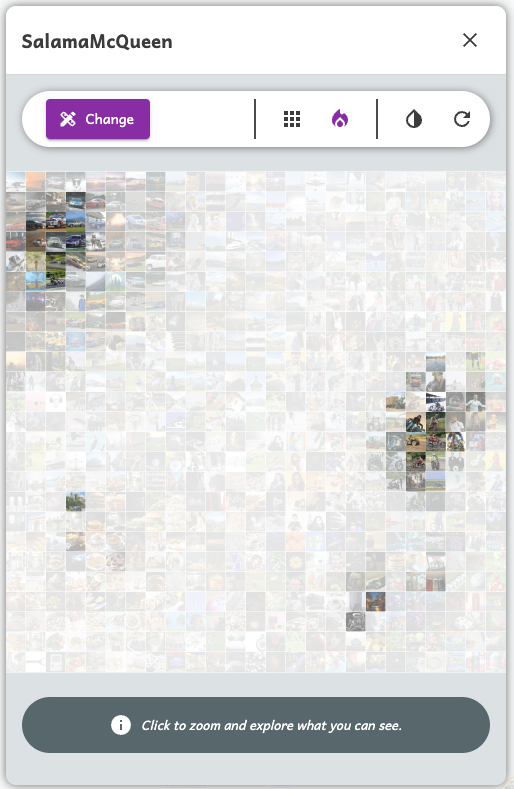}
        \caption{Users can see a map of probable recommendation space.}
        \label{fig:app_recspace}
    \end{subfigure}%
    ~ 
    \begin{subfigure}[t]{\miniminiteacherviewsize\textwidth}
        \centering
        \captionsetup{justification=centering}
        \includegraphics[height=200px]{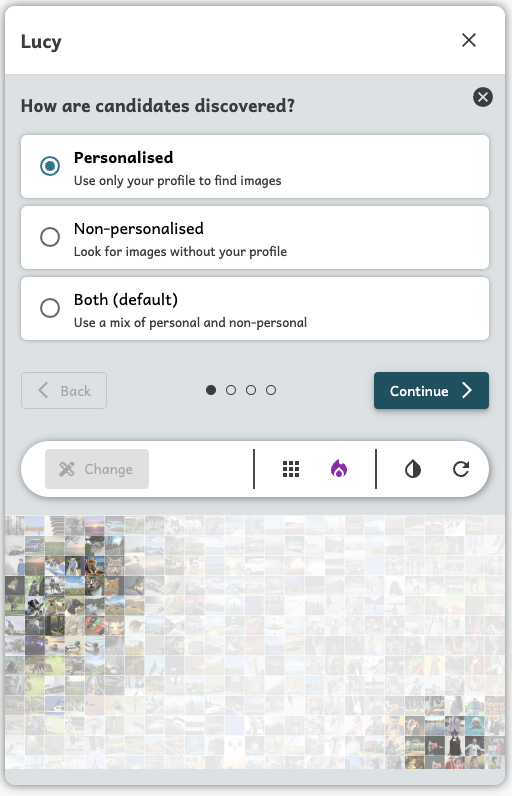}
        \caption{Users can change the basis on which recommendations are made.}
        \label{fig:app_changingrecs}
    \end{subfigure}
    ~ 
    \begin{subfigure}[t]{\miniminiteacherviewsize\textwidth}
        \centering
        \captionsetup{justification=centering}
        \includegraphics[height=200px]{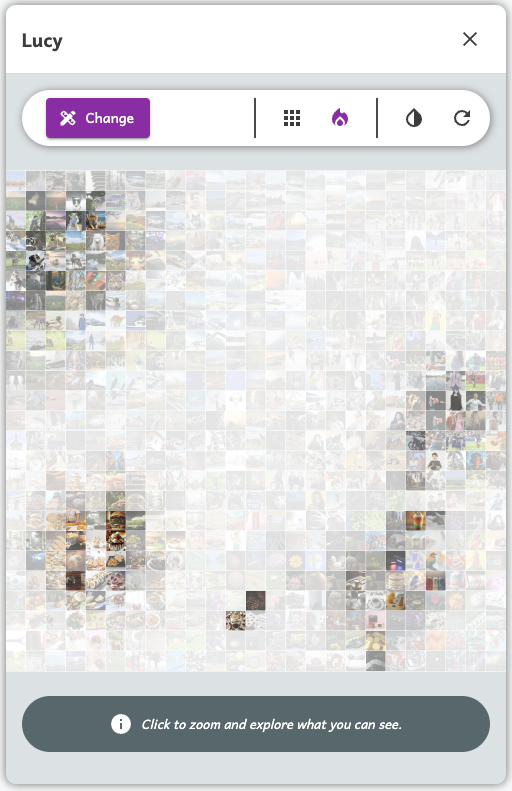}
        \caption{A heat map visualizes both what a user can and cannot see.}
        \label{fig:app_recspace2}
    \end{subfigure}%
    \caption{Somekone provides a view of each user's ``bubble,'' or the pool from which the user's recommendations are most likely drawn (Fig.\ \ref{fig:app_recspace}, \ref{fig:app_recspace2}).  User can experiment on different algorithms and approaches to making recommendations (Fig.\ \ref{fig:app_changingrecs}). 
    }
    \label{fig:recommendation_systems}
\end{figure*}

\paragraph{The recommendations view.}
In recommendations view mode, students can see the next images to appear in their feed, along with explanations for why each image is recommended (Fig. \ref{fig:app_recommendations}).  These recommendations are based on either collaborative filtering, content-based filtering, image co-engagement, random selection, or a combination of those.  The view emphasizes that the system does not need to ``understand'' the content or meaning of images but relies on user engagement data.  What is more, students can see how their browsing influences not just their own experience on the platform but impacts everyone else's experience, too.  To illustrate how recommendation systems polarize views, Somekone provides a heat map showing which images are likely to be seen or missed based on the recommendation algorithm (Figs. \ref{fig:app_recspace} and \ref{fig:app_recspace2}).  Users can experiment, in real-time, how their image feed browsing changes the content they will encounter (their ``bubble'') as the system further adjusts their profile, and they can also experiment how changing the recommendation parameters affect that bubble.

\paragraph{Recommendation algorithm parameters.}
Somekone allows students to experiment with different kinds recommendation algorithms and their parameters, visualizing their impact on their content feed in real-time.  Figure \ref{fig:app_changingrecs} shows swapping between different recommendation algorithms and parameters.  This feature prepares students for the changes in online platforms mandated by the EU's new Digital Services Act (DSA), which requires transparency about recommendation systems, especially ``the main parameters used in their recommender systems, as well as any options for the recipients of the service to modify or influence those main parameters'' (DSA, Article 27 §1).  As students adjust the parameters, they can observe their impact immediately on their content bubble, visualized as a heat map of what users can and cannot see (Figs. \ref{fig:app_recspace} and \ref{fig:app_recspace2}).

\paragraph{Classroom views of social networks.}
Somekone provides a range of visualizations for teachers to illustrate the emerging structures from students' engagement with the content.  For instance, teachers can show topic affinities, or how closely related are different topics based on topic co-engagement (Fig. \ref{fig:scr_topic_affinity}); image co-engagement, or what images are close to each other based on users' engagement with them (Fig. \ref{fig:scr_image_coengagement}); different tables of user engagement (Fig. \ref{fig:scr_table_view}); and different sizes of clusters (Fig. \ref{fig:scr_topic_clustering}).

\paragraph{Supported platforms.}
Somekone is completely browser-based and it works on browsers that fully support WebRTC data channels and ES11 (Chrome 56+, Firefox 44+, Safari 15.4+ and Edge 79+).  It has been tested with roughly 400 children in dozens of schools with devices ranging from laptops (Chromebooks, Windows, Mac and Linux laptops) to mobile phones (Android, Apple) and tablets (Android, Apple).  The number of concurrent users is recommended to be up to 30 to ensure that the tuned social network visualizations remain sufficiently detailed on the classroom projector, although larger numbers of users are possible.

\paragraph{Privacy and Security.}
Designed with children's data privacy in mind, Somekone adheres to the EU's GDPR regulations.  It does not collect, send, or store any identifiable data outside the classroom.  All data are kept local, stored only on the teacher's laptop for the duration of the session, and automatically deleted when the session ends.  The only external data retrieved are the app itself and the image and label dataset.  The tool uses WebRTC-based peer-to-peer communication, and requires only local network connection.  Children are not allowed to upload personal images to avoid introducing uncurated or age-restricted content into the classroom.

% The volume, velocity, and variety of data is key to understanding its privacy and security implications (e.g., Barassi, 2020)—but the extent of data collection is opaque to users.

\section{Enabling Research And Analytics}

\begin{figure}[!b]
 \centering
 \includegraphics[width=0.44\textwidth]{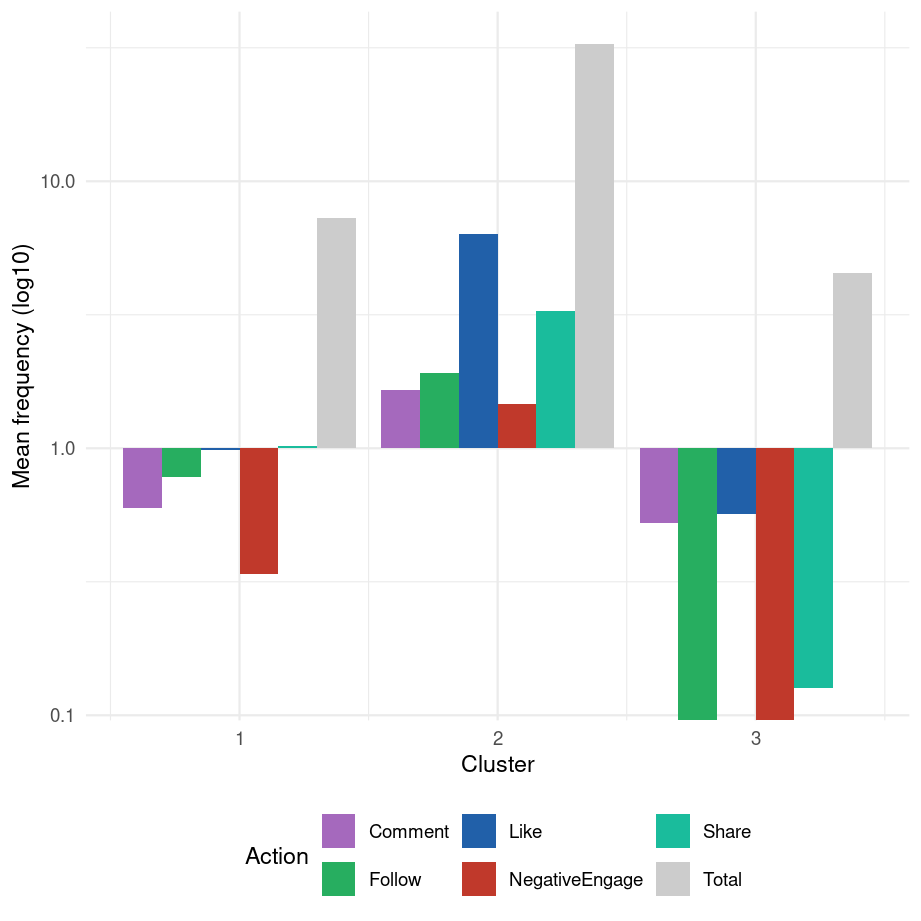} % Reduce the figure size so that it is slightly narrower than the column.
 \caption{Mean value for each variable per cluster
}
 \label{figclusters}
 \end{figure}

Everything learners do on Somekone is recorded on the teacher's machine for the duration of the session, and automatically deleted at the end of the session.  For research purposes, it is possible to save an anonymized version of log data for further analysis. The logging mechanism records fine-grained data with time-stamps which enables a wide range of analytics and most importantly, temporal analytics offering a rich understanding of what and when students do online. For instance, the logs can be used to analyze the frequencies of students interactions, create networks of their shared preferences, study the sequences of their clicks, their transitions between actions or their process of navigating the tool, to mention a few. The following example shows how temporal analytics can be used to study the sequences of students' interactions with the tool.

To study how students navigate and learn how the Somekone social media tool works, and given that students typically have different approaches to using technology, we clustered action log data of 209 learners from grades 5 and 8 in twelve schools in Finland; those data were collected from a four-hour social media education class that aimed at uncovering how social media platforms collect and use data about users (reported in Authors, forthcoming).  Two hours of that intervention used Somekone to illustrate social media mechanisms.  

The clustering was based on learners' use of Somekone using a person-centered approach. A person-centered approach groups individuals into homogeneous groups based on shared similarities. We applied latent profile analysis (LPA) using the R package \texttt{mclust} to cluster students based on the number of actions of each type performed in Somekone  \cite{Scrucca2024-xv}. We fitted 10 LPA models and chose the model with the lowest BIC, the highest entropy as well as having no cluster below 5\% according to the guidelines by Scrucca et al. \shortcite{Scrucca2024-xv}. A three-cluster model was the best solution, corresponding to three unique patterns of using Somekone. The clustering quality was confirmed by the average posterior probability value (0.98) as well as the entropy (0.98). The three clusters were labeled and described.  Figure \ref{figclusters} shows the average variables for each of the three Somekone usage profiles. Thereafter, we studied the succession of students’ interactions with Somekone using sequence analysis to understand how their behavior unfolds over time. For this purpose, we used the R package \texttt{TramineR} \cite{Gabadinho2011-lm}, specifically distribution plots for each cluster. In distribution plots, each time-point has a bar with different colors, and each color is proportional to the percentage of action at this time point.

\begin{figure*}[!t]
 \centering
 \includegraphics[width=0.99\textwidth]{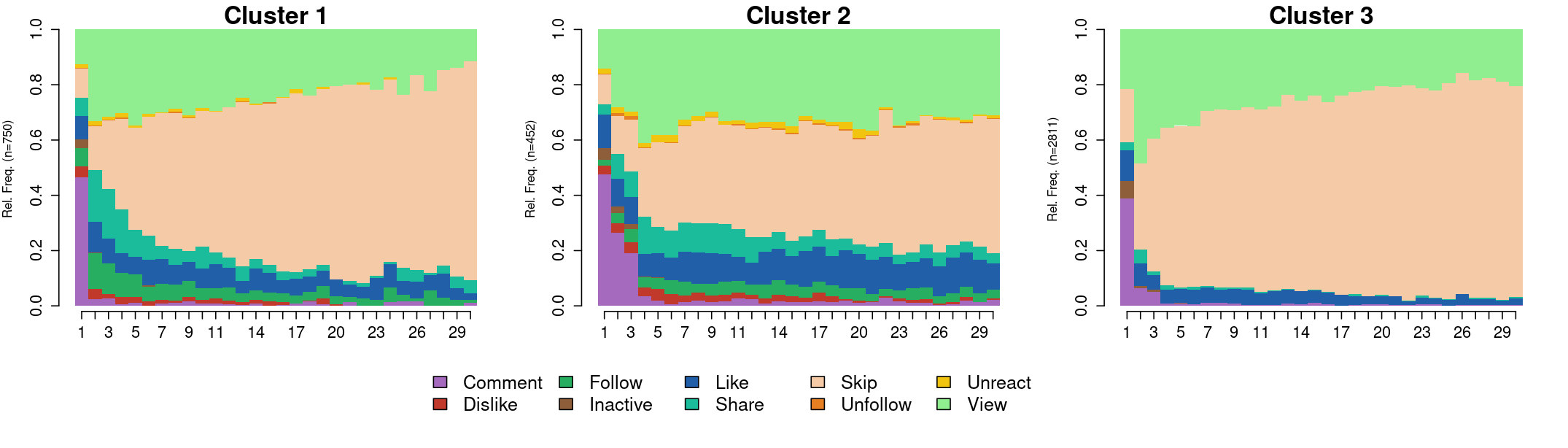} 
 % Reduce the figure size so that it is slightly narrower than the column.
 \caption{Sequence distribution plot of students' usage profile evolution}
 \label{figseq}
 \end{figure*}

\textbf{Cluster 1 - Browsers}: This group shows very low engagement with content across all interaction types, such as commenting, liking, sharing, following, and negative reactions. Their overall connection with the content is minimal, indicating they rarely interact with the materials other than browsing. 

\textbf{Cluster 2 - Engagement enthusiasts}: This group has the highest engagement with content, interacting frequently through various means—liking, sharing, following, and commenting. Their engagement is broad, active, and consistent, without focusing on any particular form of interaction.

\textbf{Cluster 3 - Selective Engagers}: This group engages with content selectively. While their overall interaction is low, they show a preference for specific actions, such as commenting and occasionally following content. They engage much less with liking and sharing, resulting in lower overall content engagement compared to Cluster 2, but they still contribute meaningfully in targeted ways.

Plotting the sequence of interactions by cluster highlights how they evolved in their interaction with the tool throughout the activity sessions (Figure \ref{figseq}). In cluster 1, we see that students are engaged initially with diverse content and then, later, most of their interactions are skipping and viewing the content with occasional engagement. In cluster 3, students exhibit initial intense engagement in commenting and liking and later, they continue engaging with only liking, skipping or viewing. A pattern that emphasizes their selectivity of their consumed content. Students in cluster 2 are intensely engaged with multiple types of content without specific focus on any of the content.

\section{Discussion}

This paper introduces Somekone, an XAI education tool for AI education that is intended to illustrate and explain the data-driven  mechanisms of social media for novice learners.  The tool focuses on four key AI concepts at a high level of abstraction---data collection (tracking), profiling, engagement, and recommendation algorithms---all in an Instagram-like familiar context.  Those high-level concepts each invoke their own, more specific AI concepts, such as clustering, engagement metrics, similarity search, and supervised learning.  The tool's main innovations are its immediacy of XAI feedback and its experiential learning environment that visualizes complex phenomena in real-time for the whole classroom, with everyone's actions affecting everyone's social media experience.  

The paper further demonstrates how Somekone action log can be exported for deeper analysis: In the example use case, latent profile analysis was used to identify three distinct types of social media users among a sample of 209 learners who participated in a pilot study in Spring 2024.  The use case shows an analysis of how the interaction patterns of those three groups evolved differently throughout the activity sessions.  The potential use cases of Somekone action logs (as a social media simulator) for social media research are a promising direction for further research and development.

The tool is also limited in a number of important ways.  Firstly, by visualizing and explaining the social media mechanisms at a high level of abstraction, Somekone is not designed to model the complexity and scale of real social media platforms.  This black-boxing and simplification is necessary for young learners, but may provide insufficient detail and depth for advanced learners.  It also risks students generalizing Somekone's simplification as a true representation of real social media, which risks misconceptions similar to ``folk theories'' \cite{eslami16,devito17}.  The Instagram interface and social media context also excludes the learning of other AI types and approaches.  

% One can also ask if the tool is a technological fix to a problem created by technology itself.  It is reasonable to ask whether unplugged social media literacy activities, critical thinking activities, or other traditional classroom activities might better serve novice learners and those in less well-resourced schools.  Along these lines, one could argue that bringing a social media simulation to school education may inadvertently normalize the use of social media at too early age.  Perhaps the issues with social media should be addressed outside the educational system rather than relying on the school system as a universal remedy for every societal problem.

Future research with Somekone will evaluate learning outcomes of AI and social media concepts at different levels, and explore stronger integration of AI ethics, privacy, and societal impacts into the learning experience.  Pilot tests have shown potential for first using Somekone to learn about AI mechanisms and then engaging students in hands-on group work to work on topics like marketing, influencing, and polarization in social media.   Teacher materials, lesson plans, and evaluation rubrics are necessary for large-scale testing.  In addition to insights to AI education, the rich action log data that Somekone generates provides possibilities for research on user behaviors, engagement patterns, and algorithmic influencing, for example.

\section*{Statement on Research Ethics}
The data presented in the use case were collected following the 2019 guidelines of the Finnish National Board on Research Integrity. Research permit was obtained from the municipal educational administration, and informed consent was obtained from the guardians of each participant as well as all participants.  All data were anonymized before use.

\section*{Acknowledgments}
Funded by the Strategic Research Council (SRC) within the Research Council of Finland (\#352859, \#352870, \#352871 and \#352876). The authors thank the January Collective.  We extend our heartfelt thanks to the teachers and children who actively participated in developing and testing this material. 

\bibliography{somekone}

\end{document}